\documentstyle[12pt,aps]{revtex}

\newcommand{\cp}{CP_{\theta \theta '}}
\newcommand{\be}{\begin{equation}}
\newcommand{\bea}{\begin{eqnarray}}
\newcommand{\ee}{\end{equation}}
\newcommand{\eea}{\end{eqnarray}}
\newcommand{\ti}{\theta \mbox{ and } \theta'}

\begin{document}
{\large
\begin{flushright}FTUV 030703 \end{flushright}
\vskip1cm
\begin{center}
{\Large \tt \bf Correlated neutral $B$ meson decays into CP eigenstates}
\vskip 1cm
\noindent {\Large E. Alvarez\footnote{e-mail: Ezequiel.Alvarez@uv.es} and J. Bernab\'eu\footnote{e-mail: Jose.Bernabeu@uv.es} \\}

\noindent {\it \small Department of Theoretical Physics and IFIC, \\
\noindent University of Valencia - CSIC, E-46100 Burjassot, Spain.}
\end{center}

\vskip 3cm
\begin{center} {\bf Abstract} \end{center}

 In the two correlated $B$ meson decay experiment we propose to measure intensities relating CP eigenstate ($J/\psi K_{S,L}$) decays on $both$ sides, which will be measurable in future upgrades of KEK and PEP.  As a CP-forbidden transition, we obtain $I(J/\psi K_S, J/\psi K_S, \Delta t) \sim \sin ^2 (2\beta)$.  We calculate in a model independent way all the possible intensities relating final CP and flavour eigenstate decays.  Under CPT-invariance, the asymmetries for processes related by CP$\Delta t$ vanish for $\Delta \Gamma =0$ and measure $\Delta \Gamma$ linearly.  We notice the impossibility to isolate the sign of $\cos (2\beta)$ without an independent knowledge of the sign of $\Delta \Gamma$.  This exhaustion of the possible Golden Plate and flavour decays provides new observables which may throw light in our present understanding of CKM physics.
\thispagestyle{empty}
\newpage
~\thispagestyle{empty}
\newpage
\setcounter{page}{1}
{\bf 1.} 
CP violation is currently the focus of a great deal of attention. It has been established in the neutral kaon system, both for mixing and decay \cite{1}, and in the neutral $B$-meson system \cite{2}.  To the measurement of $\sin  (2\beta)$ with better and better precision \cite{3}, where $\beta$ is the CP-phase between the top and charm sides of the $(b-d)$ unitarity triangle \cite{4}, recent analyses have moved to a more general framework to extract the lifetime difference and CPT - and T - violating parameters \cite{5}.  To within the experimental sensitivity, the Cabibbo-Kobayashi-Maskawa (CKM) mechanism \cite{6} of the Standard Model is verified.  In this description, all the CP-violating observables depend on a unique phase in the quark mixing matrix, going beyond each neutral meson system.  Any inconsistency between two independent determinations of the CKM-phase is an indication for new physics.  Inside the $B$-meson system, an experimental determination of the CP-phase $\gamma$, where $\gamma$ is the angle between the up and charm sides of the $(b-d)$ unitarity triangle, would be crucial in this respect.  There is at present an intense effort \cite{7} to search for the appropriate decay channels leading to a good determination of $\gamma$.  A non-vanishing value of both $\beta$ and $\gamma$ will again prove CP-violation in mixing and decay for the $B$-system.

The search of new physics in CP-violating observables is reinforced by the following two main reasons: $i)$ the dynamic generation of baryonic asymmetry in the Universe \cite{8} requires a magnitude of CP violation stronger than that provided by the Standard Model; $ii)$ essentially all extensions of the Standard Model introduce new sectors with additional sources of CP-phases, for example, in supersymmetry \cite{9}.  With an eye to these motivations, we study in the present paper the correlated decays of the neutral $B$-mesons into CP eigenstates in a general framework.  We will compare the {\bf time-dependent intensities} associated with decays to CP eigenstates {\it in both sides} with those for flavour-CP and flavour-flavour eigenstates in order to look for new characteristic effects and regularities.  Needless to say, the observation of CP-CP decays, like $(J/\psi K_S, J/\psi K_S)$, is highly demanding in statistics: upgraded facilities, like the Super $B$-Factories \cite{10}, would be needed.  This paper allows therefore to close the circle for correlated decays into flavour and CP-eigenstates in the spirit of the theoretical study of Ref. \cite{mcb2} where the CP-tag was introduced.  This motivated the experimental analysis of Ref. \cite{5}.  The present results include the effect of the lifetime difference $\Delta \Gamma$ in first order, compatible with $Re\ \epsilon \neq 0$, where $\epsilon$ is the CP-mixing in the effective Hamiltonian.  Time-integrated intensities for these decays into CP-CP eigenstates were considered by Wolfenstein \cite{12} and Gavela et al. \cite{13} as a test of the superweak model for CP violation. Time-ordered correlated decays have also been considered in Ref.\cite{14} to test CPT invariance.  Methods to extract $\Delta \Gamma$ have been considered \cite{mcb2,15} from time-dependent and time-independent measurements.  CP-tag decays have been also proposed \cite{16} to extract $\gamma$.

\vskip1cm
{\bf 2.}
In field theory a symmetry operation is only well defined if the corresponding transformation respects the whole Hamiltonian.  If the considered operation is not a good symmetry of the problem, there is no well defined transformation.  Instead, there are infinite transformations of the chosen type, none of them leaving the Hamiltonian invariant.  Regarding CP operation, the invariance requirement for strong and electromagnetic terms is not enough to completely fix the transformation, some phases still remain undetermined. Although this phase freedom is irrelevant for interactions mediated by neutral currents, it affects the charged weak currents: there is no choice of the CP operator that leaves the Hamiltonian invariant.  In any case, with no charged currents being taken into account, there will be two diagonal matrices $\ti$ for up and down quarks which remain undetermined.

As it can be easily seen once the gauge symmetry has been spontanously broken and the weak basis changed to the physical basis, the CP operators act on the physical quarks fields in the following way,
\bea 
q'_j(\vec{x},t) \stackrel{\rm CP}{\rightarrow} e^{2i\theta'_j} \gamma_0 {\cal C} \bar{q'}_j^T(-\vec{x},t) \ \ , \label{trafo1}\\
\bar{q'}_j(\vec{x},t) \stackrel{\rm CP}{\rightarrow} -{q^\prime}_j^T(-\vec{x},t){\cal C}^{-1} \gamma_0 e^{-2i\theta'_j} , \label{trafo2} 
\eea
$(j=d,s,b)$, where $\cal{C}$ is a unitary $4\times 4$ matrix satisfying the condition ${\cal C} ^{-1} \gamma_\mu {\cal C} = - \gamma_\mu^T$, usually taken to be ${\cal C}=i\gamma^2\gamma^0$.  (In the notation we follow, similar relations hold for the up sector but with no primes in the $q$'s nor in the $\theta$'s.)  
Therefore we say that the CP operator is defined up to two diagonal matrices $\theta$ and $\theta'$, each of them containing the phase angles which enter into the transformation through Eqs.$\,$(\ref{trafo1}) and (\ref{trafo2}).  Now, with this natural incorporation of $\ti$, we have a whole family of operators, $\cp$, which 
leave invariant the Hamiltonian without the charged current term, i.e.
\be
[ \cp \ ,\  H - H^{cc} ] =0, \ \ \forall (\theta, \theta^\prime) ,
\ee
where $cc$ stands for charged currents.  

If we want to extend this CP symmetry to the charged current term, due to the existence of the three family CKM matrix, there is no possible choice of $\theta$ and $\theta'$ which leaves also invariant the whole charged current Hamiltonian.  There are, however, different choices of $\ti$ which leave invariant different pieces of $H^{cc}$.  

If there is a $B$-decay into a CP-eigenstate, like the Golden Plate decay $J/\psi K_{S,L}$, let $\delta H_A$ be the specific hermitian piece of the Hamiltonian responsible of the decay.  In the case $\delta H_A$ contains a single definite weak phase, there is a choice of the CP operator $\cp\equiv CP_A$ such that $[\delta H_A, CP_A]=0$.  With a well defined $CP_A$ operator, we obtain the $CP_A$ eigenstates as
\bea
| B_{\pm _A} \rangle \equiv \frac{1}{\sqrt{2}} \left( |B^0\rangle \pm CP_A | B^0 \rangle \right) ;\ \ \ CP_A | B_{\pm _A}  \rangle = \pm | B_{\pm _A} \rangle.
\label{cadorna}
\eea
Then the decays governed by $\delta H_A$ conserve $CP_A$.

The explicit phases $\ti$ that leave $\delta H_A$ invariant for the Golden Plate decay are easily found.  We work to $O(\lambda^4)$ in the quark mixing matrix, where $\lambda$ refers to the Wolfenstein notation \cite{wolf} which incorporates the experimental hierarchy among families.  In this way, CP violation in the $K$-system is neglected and $J/\psi K_{S,L}$ are CP eigenstates.  The tree level decay amplitude for $B^0 \to J/\psi K^0$ and $\bar B^0 \to J/\psi \bar K^0$ is left invariant if $\theta^\prime_b - \theta^\prime_s$ is chosen as
\be
e^{i(\theta^\prime _b-\theta^\prime _s)} \equiv  \frac{V_{cs} V_{cb}^*}{|V_{cs}
V_{cb}^*|} .
\label{*}
\ee
Similarly, the $K$-mixing effective Hamiltonian leading to $K_{S,L}$ is left invariant for
\be
e^{i(\theta^\prime _s - \theta^\prime _d )} \equiv \left. \frac{V_{cd} V_{cs}^*}{|V_{cd} V_{cs}^*|} \right |_{O(\lambda^4)}.
\label{def1}
\ee

As a consequence, due to the cyclic relation, the $\theta$'s difference that corresponds to the $b-d$ sector becomes
\be
e^{i(\theta^\prime _b -\theta^\prime _d)}=\left. \frac{V_{cd} V_{cb}^*}{|V_{cd}
V_{cb}^*|}\right |_{O(\lambda^4)} \ .
\label{eq:cpdir}
\ee
The charm side of the $b-d$ unitarity triangle defines thus, to $O(\lambda^4)$, the $CP_A$-conserving direction, where $A$ is associated with the phase choice of Eq.$\,$(\ref{*}) \footnote{Descending to $O(\lambda^3)$, the phase choice of Eq.$\,$(\ref{*}) not only leads to $\delta H_A$ invariant but the entire Hamiltonian $H_{b-s}$ for the $b-s$ sector is left invariant, i.e., the $b-s$ unitarity triangle collapses to a line.  In that case, Eq.$\,$(\ref{eq:cpdir}) defines the CP-conserving direction \cite{mcb} to order $\lambda^3$ for the CP-violating $b-d$ triangle.} .  We notice the rephasing covariance of Eqs.$\,$(\ref{*})-(\ref{eq:cpdir}), with well defined transformation as the direction of the sides of the unitary triangles.

\vskip1cm
{\bf 3.} In a $B$ factory operating at the $\Upsilon (4S)$ peak, correlated pairs of $B$
mesons are produced.  Due to Bose statistics requirements and the instrinsic spin of the $\Upsilon(4S)$, the initial state may be written in the center of mass
as
\be
|i\rangle = \frac{1}{\sqrt{2}} \left( |B^0 (-\vec{k}), \bar{B}^0 (\vec{k}) \rangle - |\bar{B}^0 (-\vec{k}), B^0 (\vec{k}) \rangle \right),
\label{initial}
\ee
where the Hilbert space is assumed to be the direct product of two single $B$ Hilbert spaces.  Suppose that there is a first decay to $X$.  This acts as a filter in the quantum mechanical sense, such that the second $B$-meson is projected into
\be
|B_{\bar X} \rangle = \frac{1}{\sqrt{2}}\left[ \langle X|B^0\rangle \, |\bar B^0\rangle - \langle X | \bar B^0 \rangle \, |B^0\rangle \right]
\ee
which satisfies $\langle X|B_{\bar X} \rangle=0$, as required by Bose statistics.  Of course, if the decay product $X$ is flavour specific, we have a flavour tag and we can establish by the conservation law which was the complementary first state $|B_X \rangle$ decaying to $X$.  Similarly, if $X$ is a CP-eigenstate like the Golden Plate decays, we have seen in last section that the $CP_A$-eigenstates $B_{\pm A}$ are well defined.  In this case, we can re-write the correlated state $|i\rangle$ as
\be
|i\rangle = \frac{1}{\sqrt{2}} \left( |B_{-_A} (-\vec{k}), B_{+_A} (\vec{k}) \rangle - |B_{+_A} (-\vec{k}), B_{-_A} (\vec{k}) \rangle \right).
\label{dactari}
\ee
As a consequence, the prepared projected state $|B_{\bar X} \rangle$ from a first $J/\psi K_S$ decay is $|B_{+A} \rangle$, whereas it is $|B_{-A} \rangle$ for a $J/\psi K_L$ decay.  The decay provides a $CP_A$-tag.  The conservation of $CP_A$ in the decay also defines the father $B$-meson, $|B_X\rangle$, in the decay.

\vskip1cm
{\bf 4.}
As is well known, the time evolution of the $B$-meson states is governed by the mass eigenstates.  Assuming CPT invariance the mass eigenstates are written

\bea
\left (
\begin{array}{c}
B_L \\
B_H \\
\end{array}
\right )
=
\frac{1}{\sqrt{1+|\epsilon |^2}}
\left (
\begin{array}{c c}
1 & \epsilon \\
\epsilon & 1 \\
\end{array}
\right ) \left (
\begin{array}{c}
B_{+_A} \\
B_{-_A} \\
\end{array} \right )
=
\frac{1}{\sqrt{2(1+|\epsilon |^2)}}
\left (
\begin{array}{c c}
1+\epsilon & 1-\epsilon\\
1+\epsilon & -1+\epsilon \\
\end{array}
\right ) \left (
\begin{array}{c}
B^0\\
CP_A \, B^0 \\
\end{array}
\right )
\label{estados de masas}
\eea
where of course $\epsilon$ depends on the definition of $CP_A$.  The time evolution of the mass eigenstates is
\be
U(\Delta t) | B_{L,H} \rangle = e^{-i \mu_{L,H} \Delta t} | B_{L,H} \rangle ,
\ee
where $\mu_k = m_k -\frac{i}{2} \Gamma_k$.

The relationship between this $\epsilon$-rephasing invariant picture and the common used notation of the $\lambda$-parameter \cite{lambda} of mixing times decay is quite simple.  With $q/p$ defining the mixing between the rephasing-variant $B^0,\,\bar B^0$ and ${\cal A} = \langle f_{CP} | \delta H_A | B^0 \rangle$ the decay amplitude, the value of $\lambda$ is 
\be
\lambda = \frac{q}{p} \frac{\bar {\cal A}}{{\cal A}} = \mbox{ rephasing invariant quantity}.
\ee    
Taking into account the invariance of $\delta H_A$ under CP$_A$, one gets the connection
\be
\pm \lambda = \frac{1-\epsilon}{1+\epsilon}= - \sqrt{\frac{H_{21} CP_{12}}{H_{12} CP_{12}^*}}.
%un saludito a la muchachada...
\label{por la proxima chica}
\ee 
where the sign depends on the final CP-eigenstate, and $CP_{12} \equiv \langle B^0 | CP_A | \bar B^0 \rangle$.  The $\epsilon$-picture reduces the CP violation to transitions between well defined neutral $B$-meson states.

Using Eq.$\,$(\ref{por la proxima chica}), the choice of $CP_A$ in Eq.$\,$(\ref{eq:cpdir}) and the top-quark dominance in the mixing, the $\epsilon$-value in the Standard Model is
\bea
\sin (2\beta) &=& -\frac{2 Im\ \epsilon}{1+|\epsilon|^2} \label{sonia2} \\
\cos (2\beta) &=& -\frac{1-|\epsilon|^2}{1+|\epsilon|^2} , \label{rosa2}
\eea   
where $\beta$ is the well known CP-phase between the top and charm sides of the $b-d$ unitary triangle, $\beta=\mbox{arg} \left( -\frac{ V_{cd}V^*_{cb} }{V_{td}V^*_{tb}} \right) $.  The sign connection of Eq.$\,$(\ref{rosa2}) depends on the $\epsilon$ definition in Eq.$\,$(\ref{estados de masas}).

\vskip1cm
{\bf 5.} We now proceed to describe the experimental variable to be measured in a correlated $B$ meson decay, the intensity \cite{intensity}.  Under the assumption of an initial state as in Eq.$\,$(\ref{initial}) the probability density of having a decay into $X$ with momentum $-\vec{k}$ at $t=t_1$ and a decay into $Y$ with momentum $\vec{k}$ at $t=t_2 >t_1$ is $| \langle X, Y | U(t_1) \otimes U(t_2) | i \rangle |^2$; $U(t)$ being the (non-unitary) evolution operator that corresponds to a single $B$ meson.  Since the important experimental variable is $\Delta t= t_2 - t_1$ we define the so-called intensity, $I(X,Y,\Delta t) =$ $\Delta t$-variable probability density of having an $X$ decay and, after a time $\Delta t$, a $Y$ decay on the other side.  After performing a change of variables to $\Delta t$ and $t_1$ and integrating in $t_1$ it results
\be
I(X,Y,\Delta t) = \int_{0}^{\infty} d t_1 | \langle X, Y|U(t_1) \otimes U(\Delta t +t_1 ) | i \rangle |^2 .
\label{intensidad}
\ee
In order to solve this integral we reason as follows.  At $t=t_1$, when the first decay to $X$, if $X$ is a flavour or Golden Plate decay then we know that the decay comes from a $|B_X\rangle$.  The projected remaining state is $|B_{\bar X}\rangle$.  In this way the only appearance of the variable $t_1$ will be in the attenuation of the norm of the state due to the elapsed time since the creation of the $B$-mesons. Therefore we may re-write the integral in Eq.$\,$(\ref{intensidad}) and solve it as follows
\be
I(X,Y,\Delta t) =  \int_{0}^{\infty} d t_1 e^{-2\bar \Gamma t_1} |\langle X,Y| 1 \otimes U(\Delta t)| B_X,B_{\bar X} \rangle |^2 =  \frac{1}{2 \bar \Gamma} |A_X|^2 |A_Y|^2 |T_{B_Y B_{\bar X}} (\Delta t) |^2.
\label{naif}
\ee
Here $\bar \Gamma$ is the average witdh of B-mesons, and $T_{B_Y B_{\bar X}} (\Delta t) = \langle B_Y | U(\Delta t) | B_{\bar X} \rangle$ is the transition amplitude in the mixing of the corresponding states.  Notice that in writing Eq.$\,$(\ref{naif}) we have assumed a conservation law in the $Y$ decay, id est, it may be flavour or Golden Plate decay.   In this way we have transformed the integral expression of Eq.$\,$(\ref{intensidad}) in the explicit result of Eq.$\,$(\ref{naif}) in terms of a $B$-meson transition, developing a powerful tool for the calculation of intensities.

We now proceed to write down the intensities for all possible experiments concerning CP and flavour decays. In so doing, we assume CPT-invariance.  The general expression for the intensity reads ($\Delta m=m_H - m_L$ and $\Delta \Gamma=\Gamma_L - \Gamma_H$)
}
\bea
I(X,Y,\Delta t) = \frac{1}{8} \frac {e^{-\bar{\Gamma}\Delta t}}{\bar \Gamma} |A_X|^2 |A_Y|^2 \left\{ a \cosh \left( \frac{\Delta\Gamma \Delta t}{2}\right) \right. &+& \left. b \cos (\Delta m \Delta t) \right. \nonumber \\ 
&+&\left. c \sinh \left( \frac{\Delta\Gamma \Delta t}{2}\right) + d \sin (\Delta m \Delta t) \right\} .
\label{laura altea}
\eea

{\large
The coefficients are easily calculated using Eq.$\,$(\ref{naif}) once the $T_{ij}(\Delta t)$ are computed using the time evolution of the mass eigenstates.  For the sake of clarity we write the bracket expression valid up to first order in $\Delta \Gamma$.  In this expansion, one has 
\be
\frac{Re\, \epsilon}{1+|\epsilon |^2} \propto  Im (\Gamma_{12} / M_{12} )
\ee
which is proportional to $\Delta\Gamma$, so that we parametrize $\frac{Re\, \epsilon}{1+|\epsilon |^2} \equiv x \Delta\Gamma$.  On the contrary, $Im\, \epsilon$ contains $\Delta\Gamma$-independent contributions and $\Delta\Gamma$-corrections, but an explicit expansion is not needed.  In Eq.$\,$(\ref{laura altea}) we approximate $\cosh (\frac{\Delta\Gamma}{2} \Delta t) \approx 1$, $\sinh( \frac{\Delta\Gamma}{2} \Delta t) \approx  \frac{\Delta\Gamma}{2} \Delta t$, and keep only first order terms in $Re\  \epsilon$ in the coefficients $a$, $b$ and $d$, and zeroth order in the coefficient $c$.  In this way for the flavour-flavour and CP-flavour decays, they read as in Tables I and II. 

\vskip.3cm
\begin{center}
\begin{tabular*}{\columnwidth}{@{\extracolsep{\fill}}|c|c|c|c|c|}
\hline
decays & ~$(l^+,l^+)$ & ~$(l^-,l^-)$ & ~$(l^+,l^-)$ & ~$(l^-,l^+)$ \\
\hline 
transition & $(B_2, B_1)$ & $(B_1, B_2)$ & $(B_2, B_2)$ & $(B_1, B_1)$ \\
\hline
\hline
$a$ & $ 1 + 4 \frac{Re\,  \epsilon}{1+|\epsilon |^2}$ & $ 1 - 4 \frac{Re\,  \epsilon}{1+|\epsilon |^2}$ & 1 & 1 \\
\hline
$b$ & $ -\left( 1 + 4 \frac{Re\,  \epsilon}{1+|\epsilon |^2}\right)$ & $ -\left( 1 - 4 \frac{Re\,  \epsilon}{1+|\epsilon |^2} \right)$ & 1 & 1 \\
\hline
$c$ & 0&0&0&0 \\
\hline
$d$ & 0&0&0&0 \\
\hline
\end{tabular*}
\end{center}
\vskip.05cm
{\bf Table I} Correlated dilepton decays, described by flavour-flavour transitions. The notation is self evident, the first row indicates the two final decay products, the second row shows the one meson transition that happens in the mixing; the remaining four rows are the coefficients for the respective intensity, see Eq.$\,$(\ref{laura altea}). 
\vskip.3cm

The first two columns of Table I are conjugated under CP and also under T.  The corresponding Kabir asymmetry \cite{kabir} becomes $\Delta t$-independent and determined by $\frac{Re\, \epsilon}{1+|\epsilon |^2}$. This is a genuine CP (and T)-violating observable, which needs both the violation of the symmetry and the absorptive part $\Delta\Gamma \neq 0$.  The equality of the third and fourth columns of Table I is a consequence of CPT invariance, assumed in this paper.
\vskip.3cm
\begin{flushleft}
{\small
\begin{tabular*}{\columnwidth}{@{\extracolsep{\fill}}|c|c|c|c|c|}
\hline
decays &$(l^-,K_L)$ or $(K_S,l^-)$ & $(l^-,K_S)$ or $(K_L,l^-)$&$(l^+,K_L)$ or $(K_S,l^+)$&$(l^+,K_S)$ or $(K_L,l^+)$\\
\hline
transition& $(B_1, B_{+ A})$ or $(B_{+A},B_2)$ & $(B_1 , B_{-A})$ or $(B_{-A},B_2)$ & $(B_2, B_{+ A})$ or $(B_{+A},B_1)$ & $(B_2 , B_{-A})$ or $(B_{-A},B_1)$\\
\hline
\hline
$a$ & $1- \frac{2 Re\ \epsilon}{1+|\epsilon|^2}$  & $1- \frac{2 Re\ \epsilon}{1+|\epsilon|^2}$ & $1+ \frac{2 Re\ \epsilon}{1+|\epsilon|^2}$ & $1+ \frac{2 Re\ \epsilon}{1+|\epsilon|^2}$    \\
\hline
$b$ &$  \frac{2 Re\, \epsilon}{1+|\epsilon|^2} $ &$  \frac{2 Re\, \epsilon}{1+|\epsilon|^2} $&$ -\frac{2 Re\, \epsilon}{1+|\epsilon|^2} $&$ -\frac{2 Re\, \epsilon}{1+|\epsilon|^2} $  \\
\hline
$c$& $ - \frac{1-| \epsilon|^2 }{1+| \epsilon |^2}  $& $ \frac{1-| \epsilon|^2 }{1+| \epsilon |^2}  $& $ - \frac{1-| \epsilon|^2 }{1+| \epsilon |^2}  $& $ \frac{1-| \epsilon|^2 }{1+| \epsilon |^2}  $\\
\hline
$d$& $  \frac{2 Im \, \epsilon}{1+|\epsilon |^2} \left( 1- \frac{2Re\ \epsilon}{1+|\epsilon|^2} \right)$& $-\frac{2 Im \, \epsilon}{1+|\epsilon |^2} \left( 1- \frac{2Re\ \epsilon}{1+|\epsilon|^2} \right)$& $- \frac{2 Im \, \epsilon}{1+|\epsilon |^2} \left( 1+ \frac{2Re\ \epsilon}{1+|\epsilon|^2} \right)$& $ \frac{2 Im \, \epsilon}{1+|\epsilon |^2} \left( 1+ \frac{2Re\ \epsilon}{1+|\epsilon|^2} \right)$\\
\hline
\end{tabular*}
}
\end{flushleft}
\vskip.15cm
{\bf Table II}  Correlated lepton-CP decays, described by flavour-CP transitions of $B$'s.  Notation is as in Table I.

\vskip.3cm

In Table II, the two correlated decays inside each column are connected by CPT and have equal intensities as a consequence of CPT- invariance.  The comparison of the first and third columns allows to build a CP-odd asymmetry.  An alternative CP-odd asymmetry can be obtained from the comparison between the second and fourth columns.  In the limit of $\Delta \Gamma =0$, they are equivalent and correspond to the well known measurement of $\sin (2\beta)$ in the Standard Model:
\be
\left. \frac{I(l^- ,K_S , \Delta t) - I(l^+ , K_S , \Delta t)}{I(l^- ,K_S , \Delta t) + I(l^+ , K_S , \Delta t)} \right|_{\Delta\Gamma=0} = \sin (2\beta ) \sin (\Delta m \Delta t) ,
\ee
where Eq.$\,$(\ref{sonia2}) has been used.  Due to CPT invariance, these connections are also T-odd.  Id est, whereas $(l^-, K_S)$ and $(l^+, K_S)$ are connected by CP, the $(l^-, K_S)$ and $(K_L, l^+ )$ channels in the same columns are connected by T.  

The comparison of the first and second columns (or between the third and fourth ones) allows to define a $\Delta t$-asymmetry, consisting in the exchange in the order of appearance of the decay products $X$ and $Y$: as a consequence, the coefficients $c$ and $d$ change sign, whereas $a$ and $b$ remain the same.  In the exact limit $\Delta \Gamma =0$, $\Delta t$ and T operations, although connecting different processes, are found \cite{mcb2} to become equivalent, i.e., the second and third columns of Table II become equal in this limit.  The first and fourth columns (or the second and third ones) are connected by the CP$\Delta t$-operation.  They should also be equal by CPT-invariance if $\Delta \Gamma =0$.  The presence of $\Delta \Gamma \neq 0$ will induce a CP$\Delta t$ -odd asymmetry which is linear in $\Delta \Gamma$, i.e. a fake CPT-odd asymmetry induced by absorptive parts. Explicitly \cite{berni} one has 
\bea
\frac{I(K_S , l^+ , \Delta t) - I(l^- , K_S ,\Delta t)}{I(K_S , l^+ , \Delta t) + I(l^-, K_S, \Delta t)} =\frac{\Delta\Gamma}{1 + \sin(2\beta ) \sin (\Delta m \Delta t) } &\left\{ \frac{\Delta t}{2} \cos (2\beta) +   4 x \sin^2 \left( \frac{\Delta m \Delta t}{2} \right) \right. \nonumber \\
& \left. + 2 x \sin (2\beta)  \sin (\Delta m \Delta t) \right\},
\label{maite}
\eea
where Eqs.$\,$(\ref{sonia2}) and (\ref{rosa2}) have been used.

The three terms of Eq.$\,$(\ref{maite}) contain different $\Delta t$-dependence, so that a good time resolution would allow the determination of the parameters.  Notice that the asymmetry is linear in $\Delta \Gamma$.  In adition, $\cos (2\beta)$ - a quantity of high interest to remove the two-fold ambiguity in the measurement of $\beta$ - is accompanied by $\Delta\Gamma$.  We conclude that the comparison between the $(K_S, l^+)$ and $(l^-, K_S)$ channels is a good method to obtain information on $\Delta\Gamma$, due to the absence of any non-vanishing difference when $\Delta\Gamma=0$.

\pagebreak
For the case of CP-CP decays the coefficients of Eq.\,(\ref{laura altea}) read as given in Table III.

\vskip.3cm
\begin{center}
\begin{tabular*}{\columnwidth}{@{\extracolsep{\fill}}|c|c|c|c|c|}
\hline
decays & $(K_L,K_L)$ & $(K_S,K_S)$ & $(K_L,K_S)$ & $(K_S,K_L)$ \\
\hline
transition & $(B_{-A}, B_{+A})$ & $(B_{+A},B_{-A})$ & $(B_{-A},B_{-A})$ & $(B_{+A},B_{+A})$ \\
\hline  
\hline
$a$&$ \frac{4|\epsilon |^2}{(1+|\epsilon|^2)^2}$ & $ \frac{4|\epsilon|^2}{(1+|\epsilon|^2)^2}$ & $2 \frac{1+|\epsilon|^4}{(1+|\epsilon|^2)^2}$ & $2 \frac{1+|\epsilon|^4}{(1+|\epsilon|^2)^2}$ \\ 
\hline
$b$&$- \frac{4|\epsilon|^2}{(1+|\epsilon|^2)^2}$ & $-\frac{4|\epsilon|^2}{(1+|\epsilon|^2)^2}$ & $4\frac{|\epsilon|^2}{(1+|\epsilon|^2)^2} $ & $4 \frac{|\epsilon|^2 }{(1+|\epsilon|^2)^2}$ \\
\hline
$c$&0&0&$2\frac{1-|\epsilon|^2}{1+|\epsilon|^2}$&$-2\frac{1-|\epsilon|^2}{1+|\epsilon|^2}$\\
\hline
$d$&$0$&$0$&$+8\frac{Re\, \epsilon \ Im\, \epsilon}{(1+|\epsilon|^2)^2}$ & $-8\frac{Re\, \epsilon \ Im\, \epsilon}{(1+|\epsilon|^2)^2}$  \\
\hline
\end{tabular*}
\end{center}
\vskip.05cm
{\bf Table III} Correlated CP-CP decays.  The notation is as in Table I.
\vskip.3cm

The transitions of the first two columns are connected by T and, as these chanels consist of CP-eigenstates, by CPT-operations.  Hence the equality of both columns for all $\Delta t$.  Any deviation of this equality would be thus a signal of CPT-invariance violation.

The decays of the last two columns of Table III are connected by $\Delta t$ and $CP\Delta t$-operations.  The corresponding asymmetry 
\bea
\frac{I(K_S, K_L, \Delta t) - I(K_L, K_S,\Delta t )}{I( K_S,  K_L, \Delta t) + I( K_L, K_S,\Delta t)} &=& \frac{\Delta \Gamma}{\frac{1}{2}\left( 1 + \cos ^2 (2\beta ) \right)+ \frac{1}{2} \sin ^2 (2\beta ) \cos (\Delta m \, \Delta t)} \cdot \nonumber \\
&&\cdot  \left\{ \cos (2\beta) \frac{\Delta t}{2} + 2 x \sin (2\beta ) \sin (\Delta m \, \Delta t ) \right\}
\label{maitules}
\eea
is linear in $\Delta \Gamma$.  As before, $\cos(2\beta)$ shows up together with $\Delta\Gamma$.  This makes impossible to measure the sign of one factor without having an independent knowledge of the sign of the other one.  The individual intensities shown in Table III have additional dependences in $\cos(2\beta)$ as shown in the denominator of Eq.$\,$(\ref{maitules}): the $a$ coefficient of $I(K_S,K_L, \Delta t)$ goes as $(1/2) [ 1 + \cos ^2 (2\beta) ]$, which once more is unable to see the sign of $\cos(2\beta)$.  The Standard Model expectation $\cos(2\beta) > 0$ \cite{cos} leads, through Eq.$\,$(\ref{rosa2}), to a $B_L$ state with preferred CP-parity equal to CP-.

The correlated channel associated to the first, or second, column corresponds to a CP-forbidden transition for all $\Delta t$.  The decays to $(K_L ,\, K_L )$ or $(K_S , \, K_S )$ are associated with transitions of $B$-states with opposite CP.  Their intensities are thus proportional to $\sin^2(2\beta)$.  The ratio to the dilepton intensity is 
\bea
\frac{I(K_S,K_S,\Delta t)}{I(l^+,l^+,\Delta t)} = \frac{|A_{J/\psi K_S}|^4}{|A_{l^+}|^4} \frac{4|\epsilon|^2}{(1+|\epsilon|^2)^2} \left( 1 - \frac{4 Re\ \epsilon}{1+|\epsilon|^2} \right) \approx \frac{|A_{J/\psi K_S}|^4}{|A_{l^+}|^4}  \sin^2 (2\beta).
\eea
where in the last step Eq.$\,$(\ref{sonia2}) and the approximation $Re\ \epsilon \approx 0$ have been used.  Notice that there is no time dependence in this ratio.  This is not a surprise because the transition amplitude between any two orthogonal states, like $B_{+A} \rightarrow B_{-A}$ or $B_2 \rightarrow B_1$, has always the same time dependence, $e^{-i\mu_L \Delta t} - e^{-i\mu_H \Delta t}$.

We may also see in Tables I and III how, independent of the details of the correlated $B$ meson state evolution, the EPR correlation imposed by Bose statistics prevails from the $\Upsilon$ decay up to the two final decays. In effect, we see that for $\Delta t=0$ it is impossible to have the same decay at both sides, independent of the CP-properties.  
 
To conclude, we have completed the analysis of correlated neutral $B$-meson decays into flavour and CP eigenstates in terms of single $B$-meson transitions in the time interval $\Delta t$ between both decays.  This is made possible by a combination of flavour and CP-tags.  In turn, the action of the discrete symmetries CP, T and CPT is thus simply described in terms of initial and final $B$-states.  A variety of new observables have been proposed to show the consistency of the entire picture.  As an example, the correlated CP-forbidden $(J/\psi K_S,\, J/\psi K_S)$ and $(J/\psi K_L,\, J/\psi K_L)$ decays have to have equal intensities by CPT-invariance.  As apparent from their CP-forbidden character, each of them is proportional to $\sin ^2 (2\beta )$, with the same $\Delta t$ dependence as the $(l^+, \, l^+ )$ correlated decay. 

These genuine symmetries are also combined with the $\Delta t$ operation of exchange of the two final states of the correlated decay.  Although $\Delta t $-exchange and T-reversal lead to different processes, they have to have equal intensities in the limit $\Delta \Gamma =0$.  We then suggest the comparison of processes connected by CP$\Delta t$-operation in order to extract linear terms in $\Delta \Gamma$, as explicitly seen in Eqs.$\,$(\ref{maite}) and (\ref{maitules}).  In the $B_d$-system, these CP$\Delta t$-asymmetries are expected to be small.  One could envisage the use of these observables to determine $\Delta\Gamma_s$ for the $B_s$-system, with expected larger width-difference, by means of the $B$-factories operating at $\Upsilon (5S)$.

\vskip.7cm
{\bf Acknowledgements.}  We thank R. Aleksan, F. Mart\'{\i}nez-Vidal, J. Mat\'{\i}as, C. Y\`eche and W. Zhengtao for discussions on the topic of this paper.  A.E. is indebted to Fundaci\'on Antorchas (Argentina), Fundaci\'on K\'onex (Argentina) and Valencia University (Spain) for their respective grants during the research of this work.  This work is supported by the Grant FPA 2002-0612 of the Spanish Programme of Particle Physics.

}
\end{document}